# 3D VIDEO QUALITY METRIC FOR MOBILE APPLICATIONS

*Amin Banitalebi-Dehkordi[1], Student Member, IEEE, Mahsa T. Pourazad[1,2], Member, IEEE, and Panos Nasiopoulos[1], Senior Member, IEEE*

[1]Department of Electrical & Computer Engineering, University of British Columbia, Canada
[2]TELUS Communications Inc., Canada
{dehkordi, pourazad, panosn}@ece.ubc.ca

**ABSTRACT**

In this paper, we propose a new full-reference quality metric for mobile 3D content. Our method is modeled around the Human Visual System, fusing the information of both left and right channels, considering color components, the cyclopean views of the two videos and disparity. Our method is assessing the quality of 3D videos displayed on a mobile 3DTV, taking into account the effect of resolution, distance from the viewers' eyes, and dimensions of the mobile display. Performance evaluations showed that our mobile 3D quality metric monitors the degradation of quality caused by several representative types of distortion with 82% correlation with results of subjective tests, an accuracy much better than that of the state-of-the-art mobile 3D quality metric.

## 1. INTRODUCTION

Several quality metrics have been designed for 2D content, but in the case of 3D, introduction of new factors such as the scene's depth range, display size and type of technology (i.e., active or passive glasses, glasses-free auto-stereoscopic displays, etc.) makes the quality assessment of 3D content much more difficult [1]. Existing 2D quality metrics cannot be directly used to evaluate 3D quality since they do not take into account the effect of depth and the binocular properties of the human visual system (HVS). Such efforts result in low correlation with subjective tests as shown in [2], where the existing 2D quality metrics were applied on the right and left views separately, and the results were averaged over two views to evaluate the quality of the 3D picture.

In the case of quality assessment of mobile 3D content, the state-of-the-art video quality measure presented in [3], and known as PHSD, assesses 3D quality by combining 2D and 3D quality factors mainly based on the MSE (Mean Squared Error) of block structures. Information of the left and right channels is fused using the 3D-DCT transform and then a map of local block dissimilarities is generated. Weighted average of this distortion map results in the 3D quality component of PHSD. MSE between disparity maps is also taken into account. Although PHSD addresses most of the shortcomings of the 2D quality metrics well, it is still strongly dependent on MSE for block matching and measuring the similarity between block structures and between depth maps. Moreover, this metric is designed mainly for measuring compression distortions and as such its performance is only verified for compressed videos.

In this paper, we propose a new and efficient full-reference 3D quality metric for mobile 3D content. Our metric combines the quality of each of the views, the quality of cyclopean view and the quality of depth maps to measure 3D visual quality. In order to assess the quality degradation caused by 3D factors in the cyclopean view, a local quality map is extracted using the structural similarity (SSIM) index presented in [4] which is used to calculate the similarity between the reference 3D frame (left & right together) and the distorted one. The information of the left and right channels is fused using the 3D-DCT transform. The variance of the disparity map and the similarity between disparity maps are used to improve the performance of the ultimate metric. Finally, we use the Visual Information Fidelity (VIF) index [5] between each of the views and between the chroma components in order to take the 2D quality factors into account. We validated the performance of the proposed metric by subjective tests, using 4 reference and 20 modified videos and 16 subjects, following the ITU-R BT.500-11 recommendation.

The rest of this paper is organized as follows: Section 2 describes the proposed metric. Subjective tests are presented in section 3 while the results and discussions are provided in section 4. Section 5 concludes the paper.

## 2. PROPOSED 3D QUALITY METRIC

Our proposed 3D quality metric for mobile devices is inspired by our previous study presented in [6]. Our metric takes into account the quality of individual views, the quality of the cyclopean view, as well as the quality of the depth map as follows:

$$\begin{aligned}HV3D = {}& w_1 VIF(Y_R, Y_{R'}) + w_4 VIF(U_R, U_{R'}) + w_4 VIF(V_R, V_{R'}) \\ & + w_1 VIF(Y_L, Y_{L'}) + w_4 VIF(U_L, U_{L'}) + w_4 VIF(V_L, V_{L'}) \\ & + w_2 VIF(D, D')^\beta \cdot \sum_{i=1}^{N} \frac{SSIM(IDCT(XC_i), IDCT(XC_i'))}{N} \\ & + w_3 VIF(D, D')^\beta \cdot \sum_{i=1}^{N} \frac{\sigma_{d_i}^2}{N \cdot \max(\sigma_{d_j}^2 \mid j=1,2,...,N)}\end{aligned} \quad (1)$$

where $Y_R$ and $Y_{R'}$ are luma information of the reference and distorted right views respectively (similarly, indexes L and L' denote the left view contents), $U_R$ and $V_R$ are the chroma information of the reference right-view, $U_{R'}$ and $V_{R'}$ are the chroma information of the distorted right-view, $XC_i$ is the cyclopean-view model for the $i^{th}$ matching block pair in the reference 3D view, $XC_i'$ is the cyclopean-view model for the $i^{th}$ matching block pair in the distorted 3D view, IDCT stands for inverse 2D discrete cosine transform, $D$ is the depth map of the reference 3D view, $D'$ is the depth map of the distorted 3D view, $N$ is the total number of blocks in the each view, $\beta$ is a constant, $\sigma_{d_i}$ is the variance of block $i$ in the depth map of the 3D reference view and SSIM is the structural similarity, and $w_1$, $w_2$, $w_3$ and $w_4$ are weighting constants. The weighting constants are chosen so that the quality components used in our method are given different importance in order to lead to the best possible results. VIF is the visual information fidelity index which quantifies the similarity between the reference view and the distorted view using the concept of mutual information, which is widely used in information theory [5]. The quality of the cyclopean view is determined by combining the corresponding areas from the left and right views. Luma information of each view is divided into 4×4 blocks, a size that was chosen after performance evaluations showed that this size significantly reduces the complexity of our approach (i.e., search, matching and variance) while allowing us to efficiently extract local structural similarities. For each block from the left view, the most similar block in the right view is found, utilizing the available disparity information. The depth value associated to a block is the median of depth values of the pixels within the block. Note that the 4x4 block size is chosen for mobile 3D video applications. The suggested block size is also compatible with the recent video compression standards, an asset if our proposed metric is used to control compression quality instead of using common 2D quality metrics [2].

The cyclopean view is modeled once the matching blocks are detected. The objective is to fuse the information of the matching blocks in the left and right views. The 3D-DCT transform is applied to each block pair to generate two 4×4 DCT-blocks which contain the "fused" DCT coefficients. Since the human visual system is more sensitive to the low frequencies of the cyclopean view [3], we only keep the first level of coefficients and discard the other ones. The sensitivity of the human visual system to contrast is considered by deriving a 4×4 Contrast Sensitivity Function (CSF) modeling mask and applying it to the 4×4 DCT-block so that the frequencies that are of more importance to the human visual system are assigned bigger weights, as shown in the following equation.

$$XC = \sum_{i=1}^{4} \sum_{j=1}^{4} C_{i,j} X_{i,j} \quad (2)$$

where $XC$ is our cyclopean-view model for a pair of matching blocks in the right and left views, $X_{i,j}$ are the low-frequency 3D-DCT coefficients of the fused view, $i$ and $j$ are the horizontal and vertical indices of coefficients, and $C_{i,j}$ is our CSF modeling mask. This mask is created using the JPEG quantization table [7]. The 8x8 JPEG quantization table is down-sampled by a factor of two in each direction to create a 4x4 mask since this mask needs to be applied to 4×4 3D-DCT blocks. Our CSF mask is designed such that the ratio among its coefficients is inversely proportional to the ratio of the corresponding elements in the quantization table of JPEG. These elements are selected such that their average is equal to one. This guarantees that, in the case of uniform distortion distribution, the quality of each block within the distorted cyclopean view coincides with the average quality of the same view. In our implementation, the VIF index between the depth map of the distorted video and the original video's depth map is used as a scaling factor in conjunction with SSIM. This allows measuring the quality of the cyclopean view more accurately since VIF is known to predict the image fidelity with more accuracy than other 2D metrics [5]. Considering that geometrical distortions are the source of vertical parallax, which causes severe discomfort for viewers, $\beta$ in the equation (1) is empirically assigned to 0.7 (resulted from a series of subjective tests), so that more importance is given to VIF index (the range of VIF and SSIM metrics by default is between 0 to 1).

The local disparity variance is calculated over a block size area that can be fully projected onto the eye fovea when watching a mobile 3D display from a typical viewing distance. Considering this fact and also including the resolution and dimensions of the 3D cell phone used in the subjective tests, the block size for the search range and for finding the disparity variance was chosen to be 28x28 to be consistent with the fovea visual focus. The following paragraph elaborates on how to choose the block size.

As it can be observed from Fig. 1, the length of a square block (in millimeters) on the screen that can be fully projected into eye fovea is calculated as follows:

$$K = 2 \times d \times \tan(\alpha) \quad (3)$$

where α is half of the angle of the viewer's eye at the highest visual acuity. The range of 2α is between 0.5° and 2°. The sharpness of vision drops off quickly beyond this range [8]. The length of the block ($K$) can be translated in pixel units as follows:

$$k = \frac{h \times K}{H} \quad (4)$$

where $k$ is the length of the square block on screen (in pixels), $H$ is the height of the display (in [$mm$]) and $h$ is the vertical resolution of the mobile 3D display.

Substituting (3) into (4) yields:

$$k = \frac{2 \times d \times h \times \tan(\alpha)}{H} \quad (5)$$

For the case of mobile 3D displays, we used the typical distance of 30 cm from the viewer's eye to the display, vertical pixel resolution of 480 and height of 68 mm (A 3D cell phone was used for our subjective tests. Details and specifications regarding the phone are provided in section 3). We also used $\alpha=0.75^\circ$ to be within the visually acceptable fovea focus range. In summary, given the above specifications, for calculating the depth-map variance of block $i$ (i.e., $\sigma^2_{d_i}$), an outer block of 28×28 is considered such that the 4×4 block is located at its centre as shown in Fig. 2. In our case, $\sigma^2_{d_i}$ becomes:

$$\sigma^2_{d_i} = \frac{1}{28 \times 28 - 1} \sum_{k,l=1}^{28} (M_d - R_{k,l})^2 \quad (6)$$

where $M_d$ is the mean of the depth values of each 28x28 block (outer block around the $i^{th}$ 4×4 block) in the normalized reference depth map. The reference depth map has been normalized with respect to its maximum value in each frame, so that the depth values range from 0 to 1. $R_{k,l}$ is the depth value of pixel ($k$, $l$) in the outer 28x28 block within the normalized reference depth map.

The maximum of HV3D occurs when the modified video is identical to the reference video. To ensure that the HV3D index has maximum value equal to 1, we divide equation (1) by its maximum value as follows:

$$H\hat{V}3D = \frac{HV3D}{HV3D_{max}} \quad (7)$$

Since the maximum possible value of SSIM and VIF in equation (1) is unity, $HV3D_{max}$ is equal to:

$$HV3D_{max} = 2w_1 + 4w_4 + w_2 + w_3 \cdot \sum_{i=1}^{N} \frac{\sigma^2_{d_i}}{N \max\{\sigma^2_{d_j} \mid j=1,2,...,N\}} \quad (8)$$

### 3. SUBJECTIVE TESTS

In order to find the weighting constants, a series of subjective tests was performed to find the weights which result in the highest correlation with subjective tests results. The weighting constants were derived based on a series of subjective tests performed on a HD polarized display as well as a 3D cell phone. In both cases the resulted weighting constants were identical. This is mainly due to the fact that the effects of display resolution, size and distance from the display are already considered in the formulation of HV3D during the design of each quality component. Table I shows the weighting constants which yield the best performance.

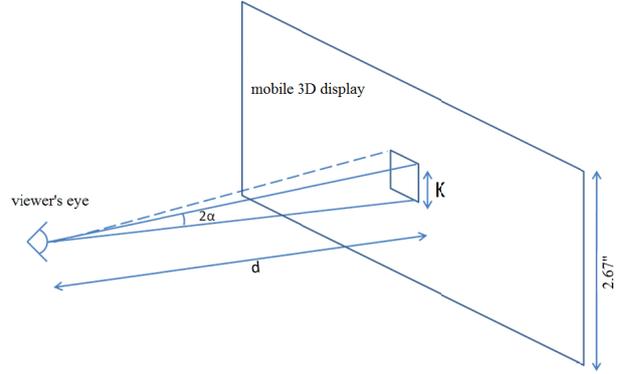

Fig. 1. Visual acuity of fovea, receptive field; relationship between the block size projected onto eye fovea and the distance of the viewer.

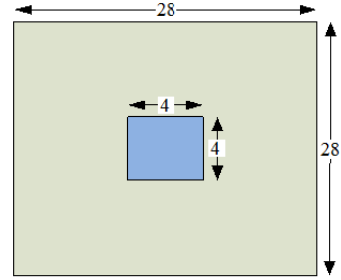

Fig. 2. Variance of disparities

To validate the performance of our proposed HV3D metric, we performed subjective tests using 3D sequences selected from the test videos in [9] and the 3D video database of the Digital Multimedia Lab (DML) at the University of British Columbia. The objective was to use 3D videos that cover a wide range of motion, brightness, and depth. The specifications of the test videos are summarized in Table II.

TABLE I: Weighting Constants

| $w_1$ | $w_2$ | $w_3$ | $w_4$ |
|---|---|---|---|
| 0.14 | 0.1208 | 0.05 | 0.1353 |

TABLE II: 3D video Dataset

| Sequence | Resolution | Frame Rate (fps) | Number of Frames |
|---|---|---|---|
| Cokeground | 480×800 (downsampled from 1080x1920) | 30 | 210 |
| Soccer2 | 480×800 (up sampled from 480x720) | 30 | 450 |
| Horse | 480×800 (up sampled from 270x480) | 30 | 140 |
| Car | 480×800 (up sampled from 270x480) | 30 | 235 |

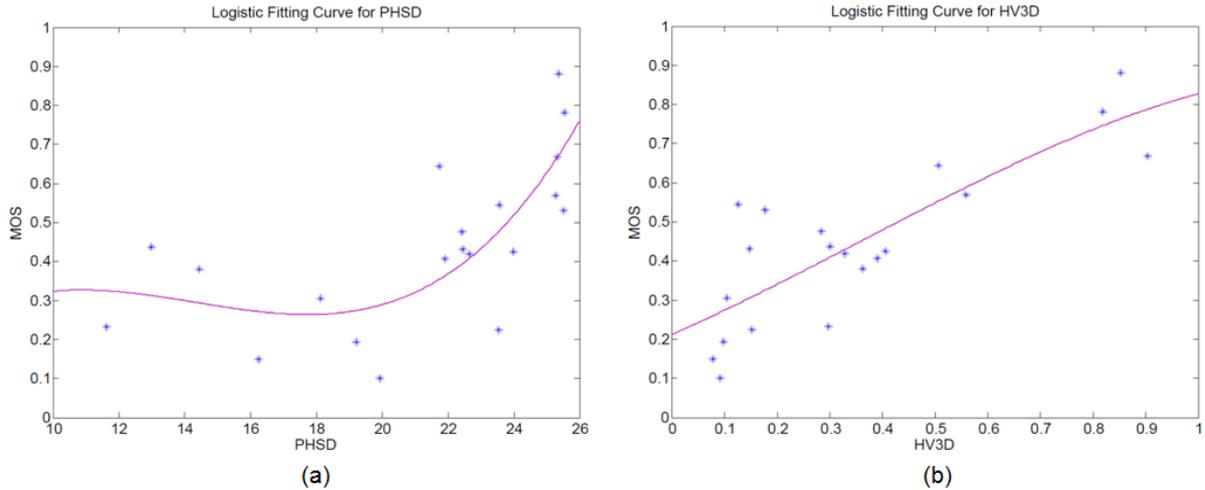

Fig. 3. Comparing the subjective results with objective results using PHSD (a) and HV3D (b) objective quality metrics

Four different types of distortions (commonly used to evaluate the performance of quality metrics) with visible impacts are applied to each of these four stereo-videos: White Gaussian noise, compression distortion (videos are encoded using the emerging HEVC standard), Gaussian low-pass filtering, and mean intensity shifting. In the case of compression, two levels of distortion were used. The quantization parameter (QP) was set at 35 and 40 to investigate the performance of our proposed metric at two different compression-distortion levels with visible artifacts. Please note that these distortions were only applied to the left & right views (not to the reference depth map). The distorted depth map was extracted from the distorted views.

Sixteen observers participated in our subjective tests, ranging from 19 to 26 years old. There was no outlier in our tests (a subject is labeled as outlier if the correlation between the MOS and the subjects' rating scores for all videos is less than 0.75). The 3D mobile device used in the tests was an LG Optimus P925G cell phone, which has 800x480 pixel resolution with display dimensions 2.67" (W) by 5.07" (H). It utilizes the parallax barrier technology to display the 3D content. Test settings were based on the MPEG recommendations for the subjective evaluation of the proposals submitted in response to the 3D Video Coding Call for Proposals [10] and the ITU-R Recommendation BT.500-11 [11]. In particular, subjects were asked to rate a combination of "naturalness", "depth impression" and "comfort" as suggested by [12].

## 4. RESULTS AND DISCUSSION

We chose to compare the performance of our quality metric with that of PHSD since this is the only quality metric that is specifically designed to measure the quality of 3D mobile content. Fig. 3 shows the logistic fitting curve for the MOS and the resulting values from each 3D quality metric for 5 different distortions. As it can be observed, HV3D outperforms PHSD, showing much stronger correlation with the MOS results.

We also used the Spearman ratio as another way of measuring the performance of our 3D quality metric. Recall that the Spearman ratio measures the statistical dependency between the subjective and objective results and assesses how well the relationship between two variables can be described using a monotonic function. The Spearman ratio for PHSD and HV3D was computed as 0.59 and 0.82, respectively, confirming the superiority of our proposed metric (23% improvement).

## 5. CONCLUSION

In this paper, we proposed a 3D quality metric customized for mobile applications. Our approach takes into account the quality of each of the views, the cyclopean view and the depth map to measure the overall perceptual 3D video quality.

Performance evaluations revealed that our method outperforms the state-of-the-art 3D quality metric for mobile content (PHSD), resulting in 82% correlation with MOS, compared to 59% achieved by PHSD.